\documentclass[prl,nofootinbib,floatfix,twocolumn,showpacs,preprintnumbers,
amsmath,amssymb,superscriptaddress]{revtex4}

\usepackage{graphicx}
\usepackage{dcolumn}
\usepackage{bm}

\def\hermes{{\sc Hermes}}

\def\pepsi{{\sc Pepsi}}
\def\jetset{{\sc Jetset}}

\begin{document}

\title{ Flavor Decomposition of the Sea Quark Helicity Distributions in
  the Nucleon \\ from Semi-inclusive Deep-inelastic
  Scattering}


\def\groupalberta{\affiliation{Department of Physics, University of Alberta, Edmonton, Alberta T6G 2J1, Canada}}
\def\groupargonne{\affiliation{Physics Division, Argonne National Laboratory, Argonne, Illinois 60439-4843, USA}}
\def\groupbari{\affiliation{Istituto Nazionale di Fisica Nucleare, Sezione di Bari, 70124 Bari, Italy}}
\def\groupcolorado{\affiliation{Nuclear Physics Laboratory, University of Colorado, Boulder, Colorado 80309-0446, USA}}
\def\groupdesy{\affiliation{DESY, Deutsches Elektronen-Synchrotron, 22603 Hamburg, Germany}}
\def\groupzeuthen{\affiliation{DESY Zeuthen, 15738 Zeuthen, Germany}}
\def\groupdubna{\affiliation{Joint Institute for Nuclear Research, 141980 Dubna, Russia}}
\def\grouperlangen{\affiliation{Physikalisches Institut, Universit\"at Erlangen-N\"urnberg, 91058 Erlangen, Germany}}
\def\groupferrara{\affiliation{Istituto Nazionale di Fisica Nucleare, Sezione di Ferrara and Dipartimento di Fisica, Universit\`a di Ferrara, 44100 Ferrara, Italy}}
\def\groupfrascati{\affiliation{Istituto Nazionale di Fisica Nucleare, Laboratori Nazionali di Frascati, 00044 Frascati, Italy}}
\def\groupfreiburg{\affiliation{Fakult\"at f\"ur Physik, Universit\"at Freiburg, 79104 Freiburg, Germany}}
\def\groupgent{\affiliation{Department of Subatomic and Radiation Physics, University of Gent, 9000 Gent, Belgium}}
\def\groupgiessen{\affiliation{Physikalisches Institut, Universit\"at Gie{\ss}en, 35392 Gie{\ss}en, Germany}}
\def\groupglasgow{\affiliation{Department of Physics and Astronomy, University of Glasgow, Glasgow G12 8QQ, United Kingdom}}
\def\groupillinois{\affiliation{Department of Physics, University of Illinois, Urbana, Illinois 61801-3080, USA}}
\def\groupmit{\affiliation{Laboratory for Nuclear Science, Massachusetts Institute of Technology, Cambridge, Massachusetts 02139, USA}}
\def\groupmichigan{\affiliation{Randall Laboratory of Physics, University of Michigan, Ann Arbor, Michigan 48109-1120, USA }}
\def\groupmoscow{\affiliation{Lebedev Physical Institute, 117924 Moscow, Russia}}
\def\groupmunich{\affiliation{Sektion Physik, Universit\"at M\"unchen, 85748 Garching, Germany}}
\def\groupnikhef{\affiliation{Nationaal Instituut voor Kernfysica en Hoge-Energiefysica (NIKHEF), 1009 DB Amsterdam, The Netherlands}}
\def\groupstpetersburg{\affiliation{Petersburg Nuclear Physics Institute, St. Petersburg, Gatchina, 188350 Russia}}
\def\groupprotvino{\affiliation{Institute for High Energy Physics, Protvino, Moscow region, 142281 Russia}}
\def\groupregensburg{\affiliation{Institut f\"ur Theoretische Physik, Universit\"at Regensburg, 93040 Regensburg, Germany}}
\def\grouprome{\affiliation{Istituto Nazionale di Fisica Nucleare, Sezione Roma 1, Gruppo Sanit\`a and Physics Laboratory, Istituto Superiore di Sanit\`a, 00161 Roma, Italy}}
\def\groupsimonfraser{\affiliation{Department of Physics, Simon Fraser University, Burnaby, British Columbia V5A 1S6, Canada}}
\def\grouptriumf{\affiliation{TRIUMF, Vancouver, British Columbia V6T 2A3, Canada}}
\def\grouptokyo{\affiliation{Department of Physics, Tokyo Institute of Technology, Tokyo 152, Japan}}
\def\groupamsterdam{\affiliation{Department of Physics and Astronomy, Vrije Universiteit, 1081 HV Amsterdam, The Netherlands}}
\def\groupwarsaw{\affiliation{Andrzej Soltan Institute for Nuclear Studies, 00-689 Warsaw, Poland}}
\def\groupyerevan{\affiliation{Yerevan Physics Institute, 375036 Yerevan, Armenia}}


\groupalberta
\groupargonne
\groupbari
\groupcolorado
\groupdesy
\groupzeuthen
\groupdubna
\grouperlangen
\groupferrara
\groupfrascati
\groupfreiburg
\groupgent
\groupgiessen
\groupglasgow
\groupillinois
\groupmit
\groupmichigan
\groupmoscow
\groupmunich
\groupnikhef
\groupstpetersburg
\groupprotvino
\groupregensburg
\grouprome
\groupsimonfraser
\grouptriumf
\grouptokyo
\groupamsterdam
\groupwarsaw
\groupyerevan


\author{A.~Airapetian}  \groupyerevan
\author{N.~Akopov}  \groupyerevan
\author{Z.~Akopov}  \groupyerevan
\author{M.~Amarian}  \groupzeuthen \groupyerevan
\author{V.V.~Ammosov}  \groupprotvino
\author{A.~Andrus}  \groupillinois
\author{E.C.~Aschenauer}  \groupzeuthen
\author{W.~Augustyniak}  \groupwarsaw
\author{R.~Avakian}  \groupyerevan
\author{A.~Avetissian}  \groupyerevan
% ...

\collaboration{The \hermes\ Collaboration} \noaffiliation

\date{\today} 
\begin{abstract}
  Double-spin asymmetries of semi-inclusive cross sections for the
  production of identified pions and kaons have been measured
  in deep-inelastic scattering of polarized positrons on a polarized
  deuterium target.
Five helicity distributions including those for three sea quark 
flavors were extracted from these data together with
re-analyzed previous data for identified pions from a hydrogen target. 
These distributions are consistent with zero for all
three sea flavors. 
A recently predicted flavor asymmetry in the polarization of 
the light quark sea appears to be disfavored by the data.
\end{abstract}

\maketitle

The quark-parton picture of nucleon structure includes the
three valence quarks that define the quantum numbers of the bound
state, gluons that mediate the strong force between the quarks, as well as
a significant presence of virtual sea quarks 
from gluon splitting and non-perturbative processes.
The relativistic motion of these bound partons 
has been thoroughly studied, mainly 
in deeply inelastic scattering (DIS) of leptons~\cite{thomas}. 
However, the photon exchange that dominates the interactions of
{\em charged} leptons limits their flavor sensitivity to 
the magnitude of the quark charge, failing to distinguish sea
quarks. The parity-violating charged-current interaction
present in neutrino scattering and W$^\pm$ production 
helps to distinguish the parton distribution functions 
(PDFs) $q(x)$ of quarks and antiquarks
of flavors $q=(up,down,strange,charm)$~\cite{pdf:cteq5,pdf:MRST2001}. 
Here $x$ is the dimensionless
Bjorken scaling variable representing the momentum fraction of
the target carried by the parton in the frame where the target has 
``infinite" momentum.

These PDFs depend on
whether the parton's helicity is equal or opposite to that
of the nucleon. The differences, or helicity distributions,
$\Delta q(x)=q^{\uparrow\uparrow}(x)-q^{\uparrow\downarrow}(x)$ 
are much less well known, not only because of the far more
limited data set for scattering of polarized charged leptons on
polarized targets, but also because polarized targets are presently
impractical with neutrino beams. A standard approach is to
further constrain the problem using the different flavor sensitivity
of hyperon $\beta$ decay data, via the additional assumption of
SU(3) flavor symmetry among the structures of the octet baryons.
The first such analysis 15 years ago~\cite{emc:g1}
revealed the celebrated ``proton spin puzzle" ---
an apparent cancellation among all the $\Delta q$'s
to make a small net contribution to the spin $1/2$
of the nucleon. Also, the strange sea polarization appeared to 
be negative. These findings have since been confirmed
with steadily improving precision~\cite{smc:NLO-g1}.

An alternative approach avoids this assumption
about SU(3) symmetry by extracting more information from the
DIS data. A quark that absorbs an energetic virtual photon 
gives rise to a ``jet" of final-state
hadrons, the composition of which reflects the flavor of the
struck quark. At the moderate energies of present fixed-target
measurements 
where few hadrons are produced,
individual
hadrons from the fragmentation of the struck quark can also serve
to ``tag" its flavor. This exploitation of hadrons detected with 
the scattered leptons in
such {\em semi-inclusive} measurements requires knowledge of the
probabilities of the various types $h$ of hadrons emerging
from a struck quark of a given flavor $q$. These probabilities are
embodied in the fragmentation functions $D_q^h(z)$, where $z\equiv
E_h/\nu$ and $\nu$ and $E_h$ are the energies 
in the target rest frame of the absorbed virtual photon 
(and hence of the struck quark) and the detected hadron.  
Although fragmentation
functions have been extracted from mostly high energy $e^+ e^-$ collider
data~\cite{frag:kkp,frag:kretzer}, their applicability at lower
energies is supported by their agreement with 
fragmentation functions~\cite{frag:klc} 
for charged pions extracted from \hermes\
measurements of hadron multiplicities, and also by the
agreement between neutral pion multiplicities measured at
widely different energies~\cite{hermes:pimult}. 

Given an adequate understanding of the fragmentation process, a
complete flavor decomposition of the quark {\em and
antiquark} helicity distributions can be extracted from sufficiently
precise measurements of double-spin asymmetries
in the cross sections for leptoproduction of various types of
hadrons. In leading order (LO), the semi-inclusive
virtual photo-absorption cross section for production of a hadron
of type $h$ takes the factorized form
\begin{equation}
    \label{eq:hadronsig}
    \sigma^h(x,z)\propto\sum_q e_q^2\ q(x)\ D_q^h(z)\,.
\end{equation}
The sum at the moderate beam energies considered here
is over quark and antiquark flavors $q=(u,\bar u,d,\bar d,s,\bar s)$, 
and $e_q$ is the quark charge.
For simplicity, we suppress here the weak
logarithmic dependence of all functions on the spatial
resolution scale corresponding to $-Q^2$, the squared four-momentum of
the exchanged virtual photon. 
The double-spin asymmetry $A_1^h$ in the above cross section is 
then given by
\begin{equation}
      A_1^h(x,z)={\sum_q e_q^2\ \Delta q(x)\ D_q^h(z)\over
      \sum_q e_q^2\ q(x)\ D_q^h(z)}\ 
        {(1+R(x))\over (1+\gamma^2)}\,,
\label{eq:hadronasym}
\end{equation} 
where the factor involving 
$\gamma^2=Q^2/\nu^2$ and $R$,
the ratio of longitudinal to transverse virtual photon cross sections, accounts
for the longitudinal component included in most parametrizations 
of the unpolarized PDFs $q(x)$. 
With asymmetry data available for a variety of hadrons
from both proton and ``neutron" targets, the
above system of equations becomes sufficiently constrained
to be solved to extract the $\Delta q(x)$ for
several flavors $q$. Two such analyses have been
reported~\cite{smc:deltaq,hermes:dq1999}, using undifferentiated hadrons. The
limited statistical precision of those asymmetry data required
severe symmetry constraints to be applied among the polarizations
of the sea quark flavors. Here we report much more precise
asymmetry data for identified pions and kaons from a
deuteron target recorded by \hermes\ in 1998--2000.
In combination with identified pions from hydrogen in 1996--1997
and with inclusive data,
they result in the first extraction of quark polarizations 
for five independent flavors  
from semi-inclusive data. Sea quarks of all three flavors are
treated independently, although in the strange
sector, only $\Delta s(x)$ could be extracted and
not $\Delta \bar{s}(x)$.

Asymmetries sensitive to the helicity
distributions require both the beam and target to be polarized.
A unique feature of the \hermes\ experiment is its target of
pure nuclear-polarized atomic gas quasi-confined in an open-ended
40\,cm long cylindrical storage cell, through which passes the $E=27.6$\,GeV
electron/positron beam of the HERA storage ring at DESY. 
The self-induced beam polarization is measured 
continuously by two independent
polarimeters using Compton backscattering of circularly polarized
laser light~\cite{hermes:tpol2,hermes:lpol}. The average beam polarization for
the deuteron (proton) data set was 0.532 (0.530) with a fractional
systematic uncertainty of 1.9\% (3.4\%).
The target cell is fed
by an atomic-beam source based on Stern-Gerlach
separation~\cite{hermes:ABS} with hyperfine transitions. 
The nuclear polarization of the atoms is flipped at 90\,s time intervals,
while both this polarization and the atomic fraction inside the target cell are
continuously measured~\cite{hermes:BRP,hermes:TGA}.  
The average value of the deuteron (proton) polarization 
was 0.844 (0.824) with a fractional systematic uncertainty of 4.4 (4.2)\%.

Scattered beam leptons and any coincident hadrons are detected by
the \hermes\ spectrometer~\cite{hermes:spectr}. Leptons are identified
with an efficiency exceeding 98\% and a hadron
contamination of less than 1\% using 
an electromagnetic calorimeter, a transition-radiation detector, a
preshower scintillation counter and a {\v C}erenkov detector. Another
unique feature of the experiment is the complete hadron 
identification provided for the deuterium data set by a
dual-radiator ring-imaging {\v C}erenkov detector~\cite{hermes:rich}.
Only pions were identified by a threshold-{\v C}erenkov detector
during the earlier hydrogen measurements.

Events were selected subject to the kinematic
requirements $Q^2>1$\,GeV$^2$, $W^2 > 10$\,GeV$^2$ and $y < 0.85$,
where 
$W$ is the invariant mass of the 
photon-nucleon system, and $y=\nu/E$.
Coincident hadrons were accepted if $0.2<z<0.8$
and $x_F\approx 2p_L/W>0.1$, where $p_L$ is the longitudinal
momentum of the hadron with respect to the virtual photon
direction in the photon-nucleon center of mass frame. 
The limit on $x_F$ suppressed most of the contamination 
of the hadron sample by target
fragmentation, as evidenced by the consistency of all the $\Delta q$'s 
extracted in $0.45<z<0.8$ with those from $0.2<z<0.45$ where
target fragmentation is more likely to appear, and by the 
negligible effect on any $\Delta q(x)$ in the range $x<0.3$ of excluding
inclusive asymmetries from the analysis.

The (semi-)inclusive virtual photo-absorption asymmetries $A^{h}_1$ 
were derived from the 
lepton scattering asymmetries $A_\parallel^{h}$ using the
relation
 $ A_1^{h}={A^{h}_\parallel/ \left[ D(1+\gamma\eta) \right]}$,
where $D$ is the depolarization factor for the virtual photon and
$\eta$ is a kinematic factor~\cite{hermes:g1p}.
The effects of internal QED radiation and instrumental resolution were 
simulated~\cite{ph:pepsi,hermes:radgen,ph:jetset}, 
and corrections were applied to $A^{h}_\parallel$ using
a technique that unfolds kinematic migration of events~\cite{hermes:miller}.
This results in small
statistical correlations between bins in $x$, which
were incorporated in the subsequent analysis.
Small corrections were applied for the combined effect of the
dependences on the hadron azimuthal angle of the spectrometer
acceptance and of the unpolarized cross section.

\begin{figure*}[t]
\includegraphics[width=\textwidth]{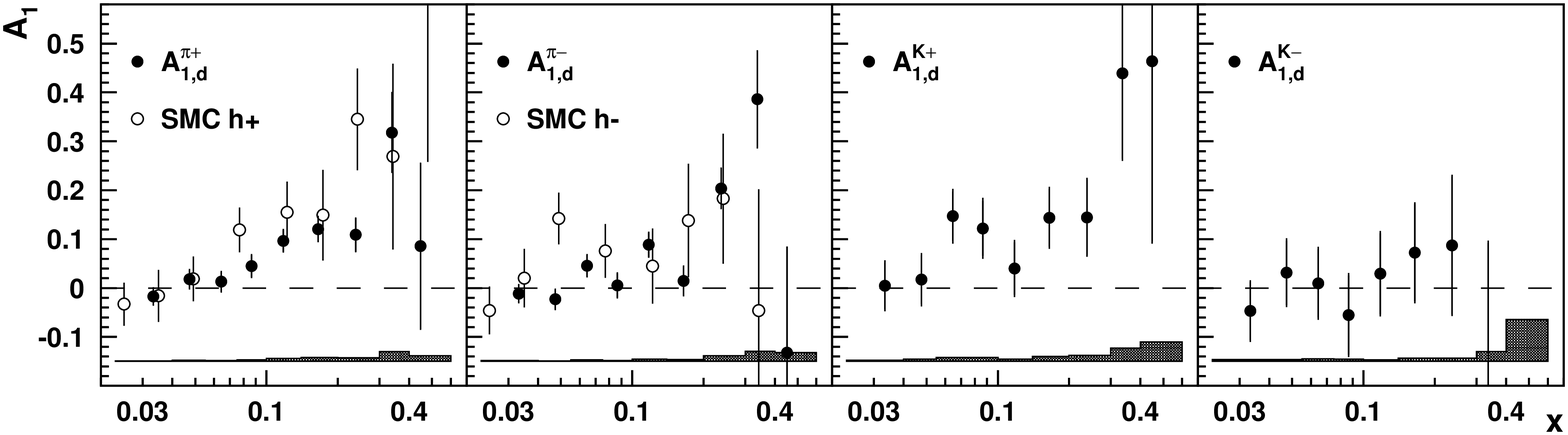}
\caption{\label{fig:A1d} Virtual photo-absorption asymmetries
$A^{h}_1$ for semi-inclusive DIS on the deuterium target as a
function of Bjorken $x$, for identified charged pions (compared to all 
charged hadrons
from SMC~\protect\cite{smc:deltaq} in the $x$-range of the present experiment),
and for identified charged kaons.
The error bars are statistical, and the bands at the
bottom represent the systematic uncertainties. 
Two data points for K$^-$ at large $x$ are off-scale
with large error bars. }
\end{figure*}

The corrected semi-inclusive asymmetries 
for the deuterium target are shown in
Fig.~\ref{fig:A1d}. They represent the first such results for
identified pions and kaons. 
The proton and inclusive deuteron asymmetries as well as 
more details about the analysis can be found in Ref.~\cite{hermes:deltaq}.
The asymmetries for the proton from 1996--1997 differ 
within the systematic uncertainties from
the published values~\cite{hermes:dq1999} due only to refinements
in the analysis such as those described above~\cite{hermes:deltaq}.
The contributions to the systematic uncertainties on $A^{h}_1$
include those from the beam and target polarizations, 
estimates of those due to 
the effects of the spectrometer acceptance,
which were studied using the \pepsi/\jetset\ DIS Monte Carlo
event generator~\cite{ph:pepsi,ph:jetset} with the photon-gluon
fusion and QCD-Compton processes enabled, 
a contribution from the ratio $R$~\cite{pdf:R1998},
and an estimate of the neglected small effect of the
spin structure function g$_2(x)$~\cite{E155:g2} representing
interference of longitudinal and transverse photons.

Integrating Eq.~(\ref{eq:hadronsig}) over $z$,  Eq.~(\ref{eq:hadronasym}) becomes
\begin{equation}
  \label{eq:hadronpuri}
  A_1^{h}(x)=\sum_q P_q^{h}(x){\Delta q(x)\over q(x)}\ 
  {(1+R(x))\over (1+\gamma^2)}\,,
\end{equation}
where $P_q^h(x)$ are the spin-independent {\em purities}:
\begin{equation}
  \label{eq:purity}
  P_q^h(x)\equiv{ e_q^2\ q(x) \int_{0.2}^{0.8} D_q^h(z)\, dz\over
    \sum_{q'} e_{q'}^2\ q'(x)\int_{0.2}^{0.8} D_{q'}^h(z)\, dz}\,.
\end{equation}
Each purity function describes the conditional probability that the
virtual photon hit a quark of flavor $q$ when a hadron of type
$h$ was detected. In the inclusive case, $P_q^h$ is replaced by 
$P_q(x)= e_q^2 q(x) / \sum_{q'} e_{q'}^2 q'(x)$.
In analogy with Eq.~(\ref{eq:purity}), the purities used in this
analysis were extracted from the above-mentioned Monte Carlo simulation 
(but now including only the effects of geometric acceptance and not
those of QED radiation and detector resolution) as
 $ P_q^h(x)={ N_q^h(x)/\sum_{q'} N_{q'}^h(x)}$,
where $N_q^h$ is the number of hadrons of type $h$, in the
geometric experimental acceptance and in the
interval $0.2<z<0.8$, that were produced when a quark of flavor
$q$ was struck. 
The simulation employs the CTEQ5L
leading order parametrization~\cite{pdf:cteq5} for the unpolarized
PDFs,
and \jetset\ fragmentation parameters that were tuned to approximate 
hadron multiplicities measured at \hermes~\cite{hermes:deltaq}.
Nuclear corrections for the deuteron target were applied, based on a
D-state probability 
$\omega_D=0.05\pm0.01$~\cite{ph:D-state:desplanques}.

Eq.~(\ref{eq:hadronpuri}) can be written in matrix form
\begin{equation}
  \vec{A}(x) = {\cal P}(x) \cdot \vec{Q}(x)\,,
\label{eq:matrixeqn}
\end{equation}
where the measured asymmetries are elements of the vector 
$\vec{A}(x)=(A_{1p},A_{1p}^{\pi^+},A_{1p}^{\pi^-},A_{1d},
A_{1d}^{\pi^+},A_{1d}^{\pi^-},A_{1d}^{\mathrm{K}^+},A_{1d}^{\mathrm{K}^-}).$ 
The matrix $\cal P$ contains the 
purities for the proton and deuteron, while the vector
$\vec{Q}(x)$ contains the quark and antiquark polarizations:
\begin{equation}
  \label{eq:flavsep}
\!\!\vec{Q}(x)=\left(
{\frac{\Delta u}{u}},\;
{\frac{\Delta d}{d}},\;
{\frac{\Delta \bar{u}}{\bar{u}}},\;
{\frac{\Delta \bar{d}}{\bar{d}}},\;
{\frac{\Delta s}{s},\frac{\Delta\bar{s}}{\bar{s}}\!\equiv 0\pm\!\frac{1}{\sqrt 3}}
\right).
\end{equation}
The set of Eq.~(\ref{eq:matrixeqn}) evaluated in all $x$-bins is solved together
for the vector $\vec Q$
by $\chi^2$-minimisation, accounting for the correlations between $x$-bins
and between the various asymmetries.
For $x>0.3$ the sea polarizations are set to zero, and the effect
on the non-sea polarizations of varying these sea polarizations
by $\pm 1/\sqrt 3$ (the standard deviation of a distribution
uniform over $\pm 1$) is included in their systematic
uncertainties.  A contribution from a similar variation of 
$[\Delta\bar{s}/\bar{s}](x)$ is applied for all $x$.

\begin{figure}[t!]
\includegraphics[width=0.9\columnwidth]{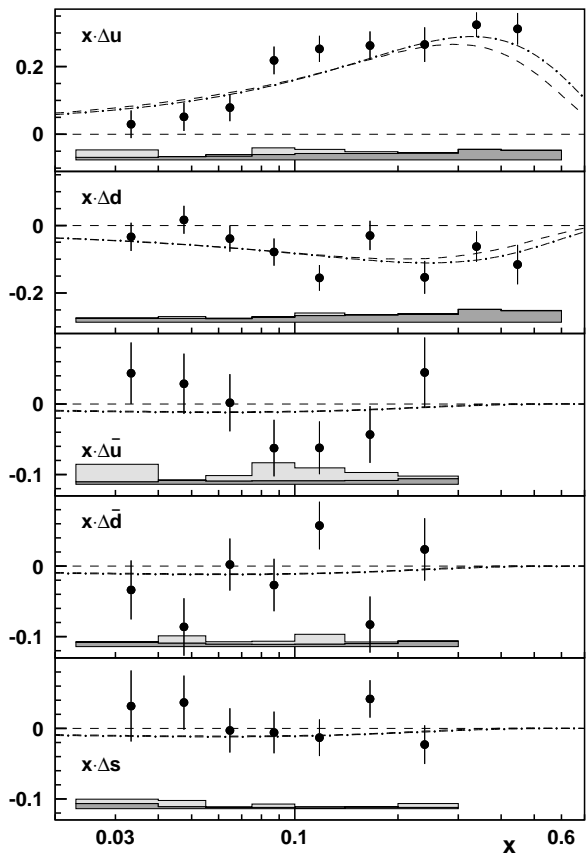}
\caption{\label{fig:deltaq} Quark helicity distributions 
at $\langle Q^2\rangle=2.5$\,GeV$^2$, as a function of Bjorken $x$, 
compared to two
LO QCD fits to previously published inclusive data shown as 
dashed~\protect\cite{pdf:grsv2000} (``standard scenario")  
and dot-dashed~\protect\cite{ph:BB} (``scenario~1") curves.
The error bars are statistical and the bands at the
bottom represent the systematic uncertainties, 
where the light area is the contribution
 due to the uncertainties of the 
    fragmentation model, and the dark area
    is the contribution due to those of the asymmetries.
}
\end{figure}

Finally, the $\Delta q(x)$ results are determined as the products
of the extracted polarizations $[\Delta q/q](x)$ and the
unpolarized PDFs from Ref.~\cite{pdf:cteq5} at the mean scale
$\langle Q^2\rangle=2.5$~GeV$^2$ of the present work. It is
assumed that the polarizations are independent of $Q^2$ within
the $Q^2$ range of this measurement. The systematic uncertainties
on the $\Delta q(x)$
include contributions from the $A_1^{h}$ data and those from 
fragmentation, which were obtained by comparing the results from two sets of
fragmentation parameters in \jetset.
Fig.~\ref{fig:deltaq} shows the results compared to two LO QCD
fits~\cite{pdf:grsv2000,ph:BB} to all available inclusive
data. These fits assume not only SU(3) flavor symmetry 
to incorporate hyperon beta decay data, but also 
explicit symmetry among the three sea quark distributions.

The extracted distributions $\Delta u(x)$ and $\Delta d(x)$ 
are consistent with previous
(semi-)inclusive results~\cite{smc:deltaq,hermes:dq1999},
but have much improved precision. The sea
distributions, extracted separately here for the first time,
are consistent with zero and with each other. 
The strange sea was previously found to be negatively polarized
in the analysis of only inclusive data assuming
SU(3) symmetry applied to hyperon beta decay data. 
However, the first moments from such analyses
evaluated over the measured $x$-range
$(\Delta s + \Delta \bar{s})/2 \equiv\int_{0.023}^{0.3} 
\Delta s(x)\,\mathrm{d}x$ are typically 
not in disagreement with that partial moment of the density extracted here:
$\Delta s = 
+0.03\pm 0.03\mathrm{(stat.)}\pm 0.01\mathrm{(syst.)}$.
\begin{figure}[t!]
\includegraphics[width=0.8\columnwidth]{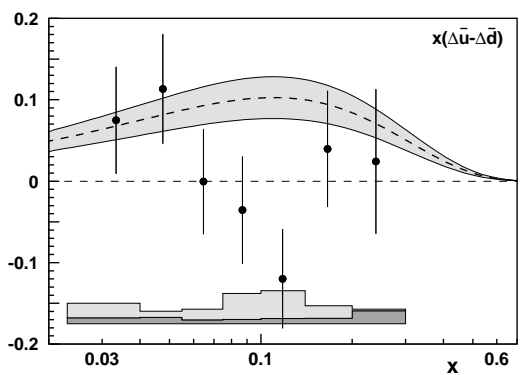}
\caption{The light quark sea flavor asymmetry 
$x\cdot(\Delta \bar{u} - \Delta\bar{d})$ 
in the helicity distributions, at $\langle Q^2\rangle=2.5$\,GeV$^2$, 
compared to 
a theoretical prediction~\protect\cite{ph:chism} (dashed curve with 
theoretical uncertainty band). 
The experimental error bars and bands have the same meaning as in 
Fig.~\protect\ref{fig:deltaq}.
}
\label{fig:DubDdb}
\end{figure}

Fig.~\ref{fig:DubDdb} shows the flavor asymmetry $\Delta
\bar{u}(x) - \Delta\bar{d}(x)$ in the light quark sea in
comparison with the prediction of a theoretical calculation based
on the chiral quark soliton model~\cite{ph:chism}. This model is
an effective field theory embodying fundamental features of QCD
that successfully describes a large body of baryon
properties~\cite{ph:chismrev}.  For example, it explains the
previously observed substantial value for the unpolarized flavour
asymmetry $\bar{u}(x) - \bar{d}(x)$~\cite{ph:chismu}. Its
prediction for the polarized moment is $\Delta \bar{u} -
\Delta\bar{d} = +0.21\pm 0.05$, to be compared to the present
experimental value of $+0.05\pm 0.06\mathrm{(stat.)}\pm
0.03\mathrm{(syst.)}$.  The difference is about $2\sigma$, when
combining all uncertainties in quadrature. Again both moments are
evaluated over only the measured range.

In conclusion, a purity-based extraction from new 
semi-inclusive DIS data has produced separated helicity distributions
for five flavors: $\Delta u$, $\Delta d$, $\Delta \bar{u}$, $\Delta \bar{d}$,
and $\Delta s$.
The polarized densities for all sea flavors are found to be consistent
with zero.

\begin{acknowledgments}
We gratefully acknowledge the DESY
management for its support,  the HERA
team for providing the polarised beams for HERMES, the collaborating
institutions for their significant effort, and our funding agencies for
financial support.

\end{acknowledgments}

\bibliography{dqletter}

\end{document}